\documentclass[aps,prl,twocolumn,amsmath,amssymb,superscriptaddress,floatfix]{revtex4}

\usepackage{amsmath}
\usepackage{amssymb}
\usepackage{amsthm}
\usepackage[pdftex]{color}
\usepackage{graphicx}% Include figure files
\usepackage{dcolumn} % Align table columns on decimal point
\usepackage{bm} % bold math
\usepackage{epic}
\usepackage{longtable}
\usepackage{ulem}   % to strike things out
\newcommand{\bs}[1]{{\boldsymbol{#1}}}

\begin{document}

\title{
Repulsive Casimir Effect with Chern insulators 
      }
      
\author{Pablo Rodriguez-Lopez} 
\affiliation{
Department of Physics and GISC, Loughborough University, Loughborough LE11 3TU, UK
            }       

\author{Adolfo G.\ Grushin} 
\affiliation{Max-Planck-Institut f\"{u}r Physik komplexer Systeme, N\"{o}thnitzer Str. 38, 01187 Dresden, Germany}
\affiliation{
Instituto de Ciencia de Materiales de Madrid, 
CSIC, Cantoblanco, E-28049 Madrid, 
Spain
            }

\date{\today}

\begin{abstract}
We theoretically predict that the Casimir force 
in vacuum between two Chern insulator
plates can be repulsive (attractive) at long distances whenever the sign of the Chern 
numbers characterizing the two plates are opposite (equal). 
A unique feature of this system is that the sign of the force can be tuned simply by turning over one of the plates 
or alternatively by electrostatic doping.
We calculate and take into account the full optical response
of the plates and argue that such repulsion
is a general phenomena for these systems as it relies on the quantized zero 
frequency Hall conductivity. We show that achieving
repulsion is possible with thin films of 
Cr-doped (Bi,Sb)$_2$Te$_3$, that were recently discovered 
to be Chern insulators with quantized Hall conductivity.
\end{abstract}

\maketitle

%\tableofcontents
%\newpage

%\medskip
%\section{Introduction}

\begin{figure*}
\begin{minipage}{0.49\linewidth}
\includegraphics[angle=0,scale=0.13,page=2]{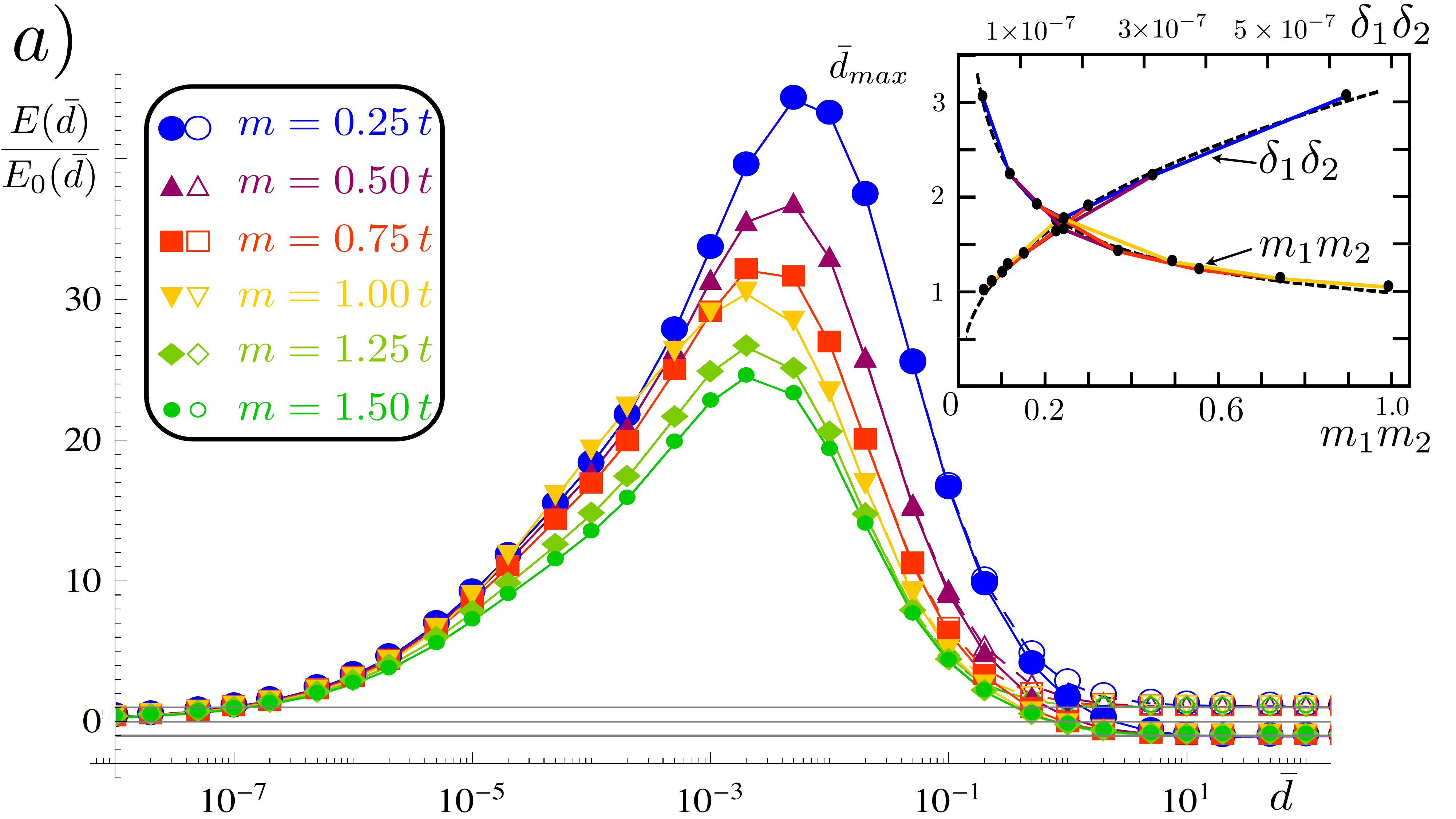}
\end{minipage}
\begin{minipage}{0.49\linewidth}
\includegraphics[angle=0,scale=0.14,page=5]{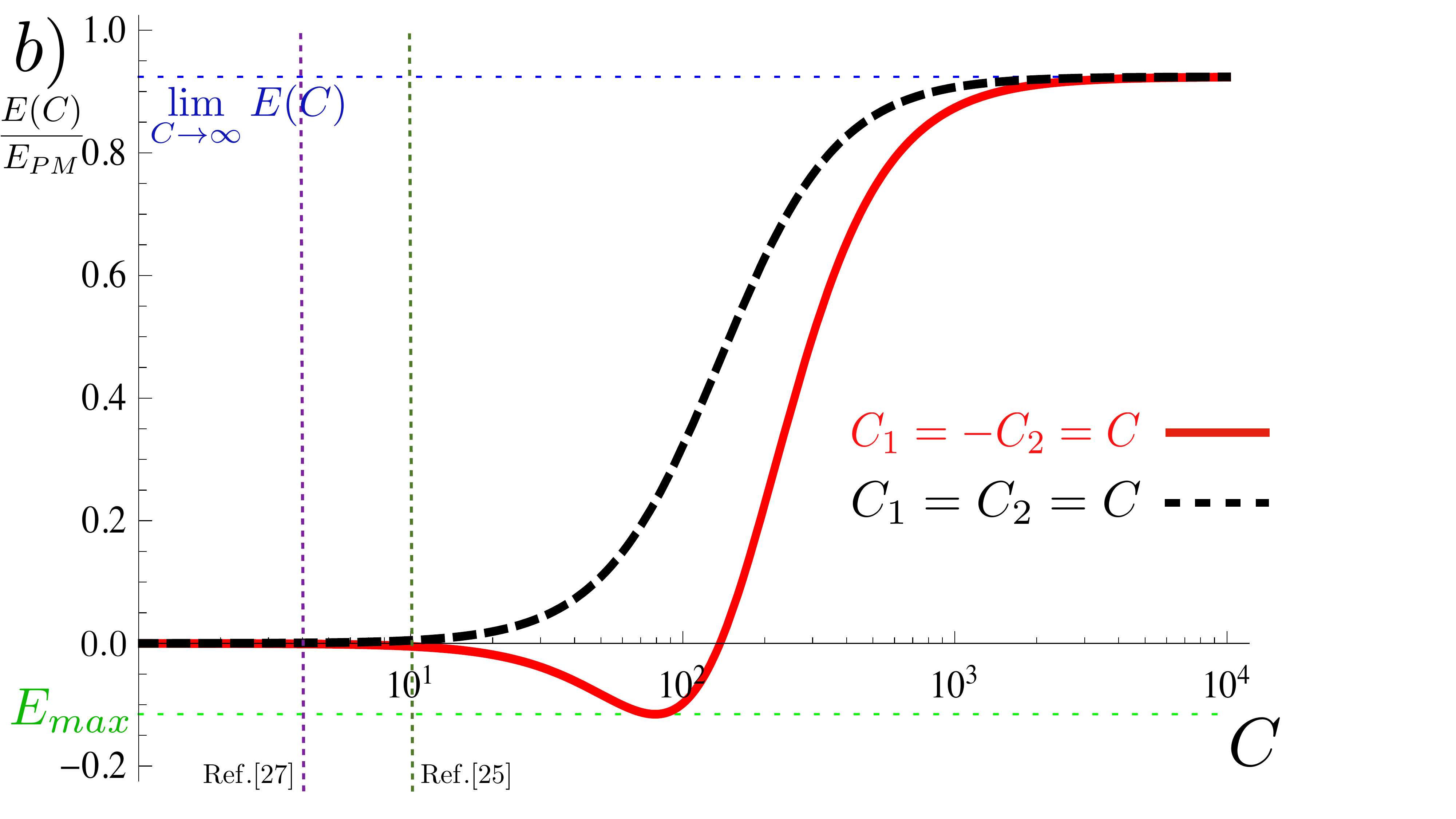}
\end{minipage}
\caption{\label{fig: mass dependence}
(Color online) (a) Casimir energy density 
$E(d)$ in units of $E_{0}(d)=\hbar c \alpha^{2}/(8\pi^2 d^{3})$
as a function of the dimensionless distance $\bar{d}=d/(\hbar c/t)$ for different values of the single particle gaps $m/t$. 
The different curves represent
the Casimir energy density for CI plates with $|C_{i}|=1$ and single particle gaps
given by $m/t\equiv m_{1}/t=m_{2}/t=\{0.25,0.5,1.0,1.25,1.5\}$. Inset: $d_{max}$ of 
$E(d)$ as a function of the gap and flatness ratio products, $m_1m_2$ and $\delta_{1}\delta_{2}$, respectively. The colors 
indicate values of $m_1$. The black-dashed line is a fit to $d_{max}\sim 0.96/(m_1m_2)^{0.41}$ and $d_{max}\sim 1019(\delta_1\delta_2)^{0.41}$. (b) Long distance limit behaviour of the Casimir energy density for two CI in units of the perfect metal result. Chern numbers to the left of the vertical dashed lines are achievable Chern numbers following \cite{FGB13} and a conservative estimate of \cite{JQL12}.}
\end{figure*}

More than half a century after its theoretical prediction, 
the Casimir effect \cite{Cas48} still stands among the most intriguing quantum phenomena. 
The relatively recent quantitative experimental access to the physics of this effect \cite{BKMM09}, 
the force experienced by objects due to quantum vacuum fluctuations, has revealed that it is still far 
from being completely understood. Despite of the development of useful calculating tools,
culminated by the development of the scattering formalism \cite{RE09,LMR06}, 
the possibility of achieving repulsion \emph{in vacuum} between two material plates is still so far unreachable experimentally. 
On the theoretical side, it is known that two dielectrics plates can repel when immersed in a medium with very specific optical properties \cite{DLP61,RK09} and that no mirror-symmetric situation can give rise to repulsion \cite{K06,B06}. These restrictions turn the search for repulsive behaviour in vacuum into a difficult challenge that can potentially solve stiction issues in device applications \cite{BKMM09,MCP09}.
Earlier proposals to achieve repulsion in vacuum include the use of magnetic materials \cite{B74}, metamaterials \cite{Rosa08,Econ09}, engineered geometries \cite{LMR10} and quantum Hall effect (QHE) systems~\cite{BV00}, where the latter was subsequently generalized to a QHE system made out of doped graphene sheets~\cite{TM12}. In \cite{GC11,GRC11} the concept of a topological Casimir effect was explored using three-dimensional topological insulators (TI) \cite{HK10,QZ11}, which owing to their topological electromagnetic response, opened the way to a tunable repulsion. Importantly, in these works, the finite frequency part of the topological response \cite{GdJ12} arising from the electromagnetic response encoded in the $\theta$-term \cite{QHZ08,EMV09} was assumed to be the quantized zero frequency response for all frequencies, which only can be justified for certain distance scales depending on material parameters.\\
In this Letter, we propose the possibility to achieve and manipulate repulsion by exploring the Casimir force arising due to the topological nature of Chern insulators (CI). These general class of two dimensional materials have a quantized Hall conductivity in the absence of external magnetic field due to the non trivial topological structure of the Bloch bands \cite{H88,HK10,QZ11}. The Chern number $C \in \mathbb{Z}$ is the topological attribute of each band that, if finite, indicates a quantized contribution to the Hall conductivity at zero frequency from each band $\sigma_{xy}(0)=Ce^2/h$. Motivated by the recent discovery of this topological phase of matter in Cr-doped TI (Bi,Sb)$_2$Te$_3$ \cite{CZF13}, in this work we show that the use of this class of materials is a feasible possibility to overcome the strict theoretical bounds to realize Casimir repulsion in realistic systems. We also show that these systems are unique in terms of controlling and reversing in a simple way the repulsive force.\\
We will first support this claim by deriving the long and short distance limit for the Casimir force 
of a generic CI lattice model. Under general assumptions, we show that whenever the sign of the Chern number characterizing the two CI plates is opposite (i.e. unequal signs of the zero frequency Hall conductivity) the system realizes Casimir repulsion at long distances. 
We further support this result by obtaining numerically the Casimir energy density for Casimir plates described by
generic CI lattice models with different Chern numbers, that can be tuned by controlling the TI thin film thickness \cite{JQL12,WLZ13,FGB13} or by achieving topological layered \cite{TB12} or multi-orbital models \cite{YGS12}. Starting from a lattice model enables us to take into account the 
\emph{complete} frequency dependence of the electronic response functions (in this case the conductivity tensor $\sigma_{ij}(\omega)$), previously overlooked \cite{GC11,GRC11,TM12},
and a key issue to ascertain any realistic Casimir force prediction \cite{RK09}. We find that the length scale
from which repulsion is achieved is inversely proportional to the products 
of the single particle gap and the Chern number of both plates. 
Finally, we show that the scaling law governing the Casimir energy density strongly 
depends on whether the Chern number of the plates is finite or zero.
Based on these results, we discuss the possibility of achieving repulsion in the recently discovered 
CI in Cr-doped (Bi,Sb)$_2$Te$_3$ \cite{CZF13,WLZ13,FGB13} and related systems.\\
The standard expression for the Casimir energy density $E(d)$ between two plates separated by a distance $d$ is \cite{DLP61,Reyn91,RE09,BKMM09}:
\begin{equation}
\label{CasimirEnergy}
\frac{E(d)}{A\hbar} = \int_0^{\infty} \hspace{-1pt} \frac{d\xi}{2\pi} \int \frac{d^2 {\bf k}_{\|}}{(2\pi)^2} \log \det \left[1 - {\bf R}_1 \cdot {\bf R}_2 e^{-2 k_z d}\right] .
\end{equation}
Here, $k_z=\sqrt{\bm{k}^{2}_{\|}+ \xi^2/c^2}$ is the wave vector perpendicular to the plates, $\bm{k}_{\|}$ is the momentum parallel to the plates and $\xi$ is the imaginary frequency $\omega = i\xi$. The $2\times2$ reflection matrices ${\bf R}_{1,2}$ contain the Fresnel coefficients
\begin{eqnarray}
\label{eq:ReflectionMatrices}
{\bf R} = \left[
\begin{array}{cc}
   R_{s,s} (i \xi, {\bf k}_{\|}) &  R_{s,p} (i \xi, {\bf k}_{\|}) \\
  R_{p,s} (i\xi, {\bf k}_{\|}) &  R_{p,p} (i \xi, {\bf k}_{\|})
\end{array} \right] .
\end{eqnarray}
The matrix elements $R_{i,j}$ $i,j=s,p$ describe parallel (perpendicular) polarization of the electric field with respect to the plane of incidence. The Casimir force per unit area on the plates is obtained by differentiating expression (\ref{CasimirEnergy}) $F=-\partial_{d}E(d)$. A positive (negative) force, corresponds to repulsion (attraction).\\
The $R_{i,j}$ components for a generic system can be computed 
by solving Maxwell's equations in the presence of the plate imposing boundary conditions 
(see Supplementary Material). For a two dimensional system
described by an optical conductivity $\sigma_{ij}(\omega)$ we find \cite{TM12} 
\begin{eqnarray}\nonumber
R_{ss} & = & - \dfrac{2\pi}{\Delta}\left( \frac{\sigma_{xx}}{\lambda} + 2\pi\left( \sigma_{xx}^{2} + \sigma_{xy}^{2} \right)\right),\\
\label{eq:fcoef}
R_{sp} & = &R_{ps}= \dfrac{2 \pi}{\Delta}\sigma_{xy} \\
\nonumber
R_{pp} & = & \dfrac{2 \pi}{\Delta}\left( \lambda\sigma_{xx} + 2\pi\left( \sigma_{xx}^{2} + \sigma_{xy}^{2} \right)\right),\\
\nonumber
\Delta & = & 1 + 2\pi\sigma_{xx}\left( \frac{1}{\lambda} + \lambda \right) + 4\pi^{2}\left( \sigma_{xx}^{2} + \sigma_{xy}^{2} \right),
\end{eqnarray}
with $\lambda={k}_{z}/\omega$ in units where $c=1$. In our convention, these coefficients are consistent with earlier results for 3D-TI \cite{Obu05,CY09,GC11} and can be related to those of Ref.~\cite{TM12} by a basis transformation.\\
To evaluate \eqref{CasimirEnergy} we employ the generic model used in Ref.~\cite{GNC12} for the CI plates. This family of two-band models captures the characteristic low energy features of any CI, (i) a quantized DC Hall conductivity
$\sigma_{xy}(0)=Ce^2/h$, where $C$ is a quantized topological integer, the Chern number of the lower band~\cite{H88,HK10,QZ11} and (ii) the insulating behaviour $\sigma_{xx}(0)=0$. The model also naturally takes into account the effect of a finite bandwidth since it is defined from a two band tight-binding model for fermions on 
a two-dimensional square lattice, with Hamiltonian 
%\begin{subequations}
\begin{eqnarray}\label{eq:Modelmain}
&&
H^{\,}_{0} =
\sum_{\bs{k}\in\mathrm{BZ}}
c^{\dagger}_{\bs{k}}
\,\bs{d}^{\,}_{\bs{k}}\cdot\bs{\sigma}\,
c^{\,}_{\bs{k}},
\\
\nonumber
&&
d^{\,}_{\bs{k};1}+\mathrm{i}\,d^{\,}_{\bs{k};2}=
t(\sin\,k^{\,}_{1}+\mathrm{i}\,\sin \, k^{\,}_{2}),
\\
&&
\nonumber
d^{\,}_{\bs{k};3}=
h^{\,}_{1}\cos\,k^{\,}_{1}
+h^{\,}_{2}\cos \, k^{\,}_{2}
+h^{\,}_{3}
\nonumber
\\
&&
\hphantom{d^{\,}_{\bs{k};3}=}
+
h^{\,}_{4}
\left[
\cos(k^{\,}_{1}+k^{\,}_{2})+\cos(k^{\,}_{1}-k^{\,}_{2})
\right],
\nonumber
\end{eqnarray}
%\end{subequations}
where 
$
c^{\dag}_{\bs{k}}\equiv
(c^{\dag}_{\bs{k},\uparrow},c^{\dag}_{\bs{k},\downarrow})
$ 
and 
$c^{\dag}_{\bs{k},s}$ 
creates a fermion at momentum $\bs{k}$ in the Brillouin zone 
(BZ) 
with $s=\uparrow, \downarrow$ being the spin or sublattice degree of freedom
and $\bs{\sigma}=(\sigma^{\,}_{1},\sigma^{\,}_{2},\sigma^{\,}_{3})$ 
are the Pauli matrices. The hopping parameters $t$ and $h^{\,}_\mu,\ \mu=1,\cdots,4$, are real and can be determined by optical spectroscopy.
This model has, at low energies, four gapped Dirac fermions that contribute $\pm 1/2$ to the total Chern number (see Supplementary Material).
In that way, tuning $h_{\mu}$ leads to different CI with different sizes of the single particle gap and Chern numbers. Thus, the Chern number of each CI plate can be chosen to take the values $C=\{0,\pm1,\pm2\}$ for generic single particle gap sizes $m$.\\
Using the Kubo formula we have calculated $\sigma_{ij}(\omega)$ for the model \eqref{eq:Modelmain} at all frequencies 
and then used the Kramers-Kronig relations to find 
$\sigma_{ij}(\omega=i\xi)$ exactly. We have also checked that the latter is equivalent to evaluating the Kubo formula at imaginary frequencies. The results for a representative case together with a typical band structure are shown in Fig.~\ref{fig: condmain} a)-c) (see the Supplementary Material for details).
\begin{figure*}
\begin{minipage}{0.49\linewidth}
\includegraphics[angle=0,scale=0.13,page=3]{cond.pdf}
\end{minipage}
\begin{minipage}{0.49\linewidth}
\includegraphics[angle=0,scale=0.14,page=1]{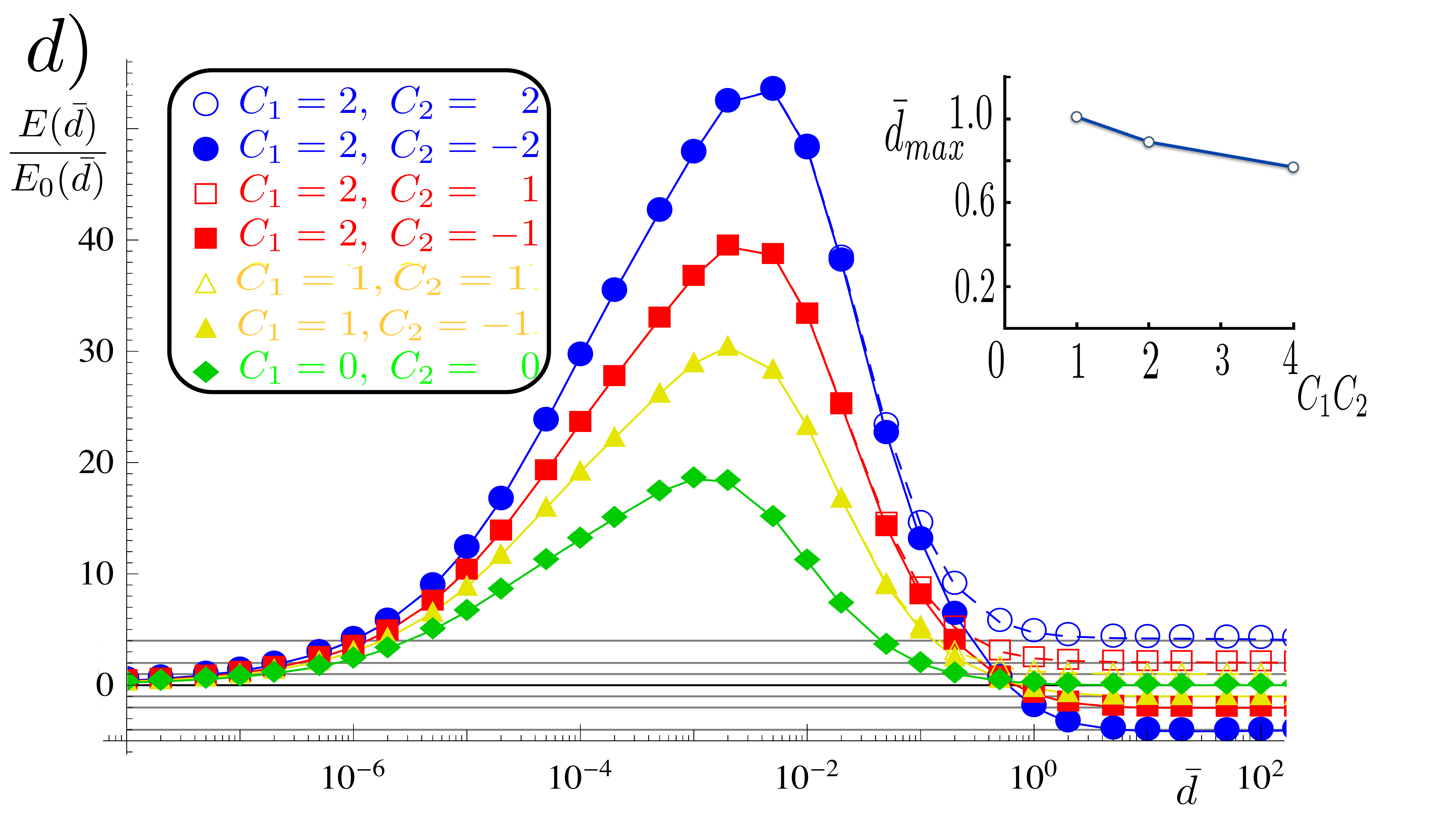}
\end{minipage}
\caption{\label{fig: condmain}
(Color online) a) Band structure, and real part of b) $\sigma_{xy}(\omega)$ and c) $\sigma_{xx}(\xi)$ as a function of real and imaginary frequencies for the model \eqref{eq:Modelmain} calculated for $h_\mu=(1,1,0.25,0)t$. The bands carry Chern number $C=\pm1$ and the conductivities are given in units of $e^2/h$. A comparison is shown between the numeric and the analytical formulas known for Dirac fermions (see Supplementary Material for details). d) Casimir energy density 
$E(d)$ in units of $E_{0}(d)=\hbar c \alpha^{2}/(8\pi^2 d^{3})$
as a function of the dimensionless distance $\bar{d}=d/(\hbar c/t)$. 
%For large distances, this ratio reaches the
%limit $C_1C_2$, and shows that the energy can be attractive (repulsive) if the product of the CI Chern numbers
%satisfies $C_1C_2>0$ ($C_1C_2<0$). 
The parameters chosen for a CI with $C=\{0,\pm1,\pm2\}$ are 
$h_{\mu}/t=\{(0,0,1,0),(1,1,\pm 1,0),(0,0,0,\pm 1)\}$ respectively, all corresponding to $m/t=1$. Inset: $d_{max}$ of 
$E(d)$ as a function of $C_1C_2$. }
\end{figure*}
The complete tight-binding calculation of the optical signatures of a lattice model of a CI is the first result of this work and will be now used to numerically evaluate \eqref{CasimirEnergy}. \\
Before proceeding, it is possible to predict the behaviour of the Casimir effect for a generic Casimir system
built up of two CI with Chern numbers $C_{1,2}$ from the asymptotic properties of $\sigma_{ij}(\omega)$. 
At short distances (large frequencies), these materials behave as ordinary dielectrics which implies attraction \cite{K06,B06}. 
For low frequencies (long distances) however, the longitudinal conductivity vanishes, since CI are insulators
and we are left only with a quantized $\sigma^{(i)}_{xy}=C_{i}e^2/h$ for each plate. 
Introducing these into \eqref{CasimirEnergy} we obtain (see Supplementary Material for details)
%
%\begin{equation}\label{repulsionlarge}
% \left. E(d)\right|_{d\to\infty} = - \frac{\hbar c \alpha^{2}}{8\pi^2 d^{3}}C_{1} C_{2},
%\end{equation}
\begin{eqnarray}\label{repulsionlarge}\nonumber
E(d)&=&-\frac{\hbar c \alpha^{2}}{8\pi^{2} d^{3}}C_{1} C_{2}- \frac{9\hbar c\,\alpha^{2}}{10\,d^{5}}b_{1}b_{2}.\\
&-& \frac{\hbar c\,\alpha^{3}}{4\pi d^{4}}\left(C_{1}^{2}b_{2}
+ C_{2}^{2}b_{1} - 2C_{1}C_{2}(b_{1} + b_{2})\right),
\end{eqnarray}
($\alpha=e^2/\hbar c$) which is valid for distances larger than the length scale set by $1/b_{i}\sim 2m_{i}$, the single particle gap, and for $C_{i}$ of order one. 
For two CI plates with opposite Chern numbers this result implies repulsion at long distances. The strength of this statement is that
it does not depend on the specific model of CI (in particular the number of bands, the concrete material realization etc),
since it only relies on the quantized Hall conductivity 
and insulator properties, which are present given that the material is a CI. 
The key issue is therefore to understand at what distances we can expect 
such a repulsive behaviour in real materials to exist.\\
To answer this question it is necessary to characterize precisely the crossover 
from repulsive to attractive behaviour and thus we have numerically computed the Casimir energy density for the model
\eqref{eq:Modelmain}. For convenience, we use the dimensionless distance $\bar{d}=d/(\hbar c/t)$ where $t$ is the hopping (typically $t\sim 1$eV and $\bar{d}\sim 1$; $d\sim 0.2\mu $m). In this calculation we include the complete $\sigma_{ij}(\omega)$ calculated numerically from the Kubo formula. %, in contrast to past approaches \cite{GC11,GRC11,TM12}.
In Fig.~\ref{fig: condmain} d) we present the Casimir energy density scaled by $E_{0}(d)\equiv\hbar c \alpha^{2}/(8\pi^2 d^{3})$ as a function $\bar{d}$ between two CI plates characterized by Chern numbers $C_{1,2}$. To unravel the effect of changing the Chern number we have chosen for this case the parameters $h_{\mu}$ such that both CI plates have the \emph{same} single particle gap $m/t=1$. For all cases we obtain repulsion (attraction) at long distances and attraction at short distances as long as $C_{1}C_{2}<0$ $(C_{1}C_{2}>0)$. 
All curves recover the analytic result \eqref{repulsionlarge} at long distances. Note also that the
Casimir force is strongly suppressed when one of the Chern numbers is zero.
The effect of changing the single particle gap $m$ while keeping $C_{1}=C_{2}$ constant is shown in Fig. \ref{fig: mass dependence} a).
In this case, we observe the same crossover behaviour as long as the Chern numbers have opposite signs.\\
 %
% \begin{figure}
%\includegraphics[angle=0,scale=0.15,page=1]{cond.pdf}
%\caption{\label{fig: Cs}
%(Color online) Casimir energy density 
%$E(d)$ in units of $E_{0}(d)=\hbar c \alpha^{2}/(2\bar{d}^{3})$
%as a function of the dimensionless distance $\bar{d}=d/(hc/t)$. For large distances, this ratio reaches the
%limit $C_1C_2$, and shows that the energy can be attractive (repulsive) if the product of the CI Chern numbers
%satisfies $C_1C_2>0$ ($C_1C_2<0$). The parameters chosen for a CI with $C=\{0,\pm1,\pm2\}$ are, in units of $t$, 
%$h_{\mu}=\{(0,0,1,0),(1,1,\pm 1,0),(0,0,0,\pm 1)\}$ respectively, all corresponding to $m/t=1$. The inset shows the decrease of the maximum of $E({\bar{d}})$ with the product $C_1C_2$.}
%\end{figure}
%
In order to optimize possible experimental systems discussed below, we now address the question on the dependence of the position maximum $d_{max}$ of $E(d)$ with the different parameters. The point $d_{max}$ results from the interplay between $\sigma_{xy}(\omega)$ and $\sigma_{xx}(\omega)$ in \eqref{eq:fcoef}. By expanding both for $\omega/t\ll1$ it is simple to estimate from \eqref{eq:fcoef} that $d_{max}\sim 1/\sqrt{|C_{1}C_{2}|m_{1}m_{2}}$ with a coefficient of order one and as long both $C_{i}\neq 0$ (see Supplementary Material). The numerical evidence for this qualitative behaviour is shown in the insets of Figs. \ref{fig: condmain} d) and \ref{fig: mass dependence} a). The former shows that $d_{max}$ indeed decreases with $|C_{1}C_{2}|$. Note that for the model \eqref{eq:Modelmain} each Chern number, when finite, can only take the values $C_{i}=\{\pm 1,\pm 2\}$ and so $|C_{1}C_{2}|=\{1,2,4\}$ providing very few points to guarantee a good fit for the power law behaviour discussed above. It is therefore more useful to study the change of $d_{max}$ against the product of the two single particle gaps $m_{1}m_{2}$ which can be tuned easily by modifying the vector $h_{\mu}$. The results are shown in the inset of Fig. \ref{fig: mass dependence} a). We find that the best fit to $d_{max}=\frac{\alpha}{(m_1m_2)^{\beta}}$ is achieved for $\beta=0.41$ and $\alpha=0.96$ for $C_{1}=-C_{2}=1$ providing evidence in favor of the simple relation above. Small deviations originate from higher values of $m_{1}m_{2}$ that might not follow this simple law. We present also $d_{max}$ as a function of the product of flatness ratios $\delta_{1}\delta_{2}$ with $\delta_{i}\equiv W_{i}/2m_{i}$ where $W_{i}$ is the band width of the filled band. \\
Complementary to the transition to repulsive behavior discussed above, we find that there are other 
intrinsic differences
between the Casimir effect between CI plates with zero Chern number and finite Chern number. Note that the leading contribution \eqref{repulsionlarge} vanishes if either or both $C_{i}=0$. From the first non-zero contribution to the long distance limit of \eqref{CasimirEnergy} we find that if
either one (both) of the Chern numbers is (are) zero the Casimir scales analytically as $\sim 1/d^{4}$ ($\sim 1/d^{5}$). Therefore, fixing a finite value for $d$ but changing from a configuration with $C_{1}=C_{2}$ to one with either or both $C_{i}=0$ will also reveal the effect of a finite Chern number. On the opposite short distance limit, we find analytically that the power law follows $\sim 1/d^{5/2}$ and independent of the Chern numbers. For both long and short distance limits, the analytical and numerical calculations agree both quantitatively and qualitatively (see Supplementary Material).\\
One of the most promising candidates to realize this effect is the recently discovered CI phase in Cr-doped (Bi,Sb)$_2$Se$_3$ \cite{CZF13}. It was experimentally shown that this material has $\sigma_{xy}(0)=e^2/h$ and $\sigma_{xx}(0)\sim 0$. Thus, a CI model such as \eqref{eq:Modelmain} captures the low energy properties since the chemical potential can be tuned to lie inside the single particle gap with a gate voltage \cite{CZF13}. 
Typical experimental values for measurable Casimir pressures and distances are pN$/m^2$ and $\mu$m respectively \cite{BKMM09}. 
For a CI of the type discovered in Ref.~\onlinecite{CZF13} the single particle gaps are of the order of $\sim 0.02$ eV \cite{LZ13} and Chern numbers up to $|C|=4$ \cite{WLZ13,FGB13} or even $|C|\gtrsim 10$ \cite{JQL12} could be reached in the thin film set up. Note that from \eqref{repulsionlarge}, increasing $C$ can result in stronger forces. However, it is instructive to take into account that, as shown in Fig.~\ref{fig: mass dependence} b), the behaviour for sufficiently high Chern numbers, beyond the validity of \eqref{repulsionlarge} can be different since (i) there exists an optimal Chern number $C_{max}\simeq 1/(\sqrt{3}\alpha)$ for which the repulsive Casimir energy is maximum reaching $\sim 10\%$ of the value for perfect metallic plates and (ii) the force turns attractive beyond $C_{0} \simeq 1/\alpha$.\\
Combining together our results we now establish an estimate for the physical realization of the effect. For two CI plates with Chern number $C\sim 10$ \cite{JQL12}  and single particle gap $m_{i}=50$ meV, the crossover lies at a distance of $d_{max}\sim 0.39~\mu$m. At the vicinity of the maximum, the typical magnitude of the pressure is a factor $10^{-2}$ smaller that of the metal-metal Casimir pressure (calculated from $E_{m}(d)/(\hbar A)=-\hbar c\pi^2/(720d^3)$) and one order of magnitude bigger than that of graphene \cite{TM12}. Although close to experimental limits, the resulting Casimir pressure at such a separation is still within observable bounds \cite{IKJ13}. Alternatively, multiorbital \cite{YGS12}  or multilayer  materials \cite{JQL12,TB12} with possible larger gaps can bring the force even further within measurable values.\\
We finish with some general remarks and consequences of the presented findings. Firstly, note that repulsion is determined by the relative chirality of the edge states of each plate, which also establish the sign of the off-diagonal Fresnel coefficients. Since the Hall effect is not induced externally, turning over one of the plates (i.e. pointing its normal in the opposite direction $\hat{n}\parallel \hat{z}\to \hat{n}\parallel -\hat{z}$) will then change the sign of the off-diagonal Fresnel coefficients. This is equivalent to reversing the sign of one of the Chern numbers and hence can turn attraction into repulsion and viceversa. This is an exclusive and differentiating feature of CI as compared to QHE systems arising from external magnetic fields \cite{TM12}, and endows the CI system with a remarkably simple way of manipulating the sign of the force. Secondly, our discussion was restricted to CI that have the chemical potential $\mu$ within the single particle gap. If this is not so, the two systems will be metallic but still have a finite Hall conductivity given by $\sigma_{xy}=\pm C\frac{m}{|\mu|}\frac{e^2}{h}$\cite{H04}. In this case, the force would be attractive due to the dominant Fermi surface contribution of $\sigma_{xx}(0)$ and $10^3$ times larger than for the insulating case. Thus, also by doping electrostatically one of the two plates it is possible to tune from attraction to repulsion and viceversa.  In addition, in Ref. \cite{GC11} 
the repulsive force between 3D-TI at \emph{short distances} resulted from the competition between the 
$\theta-$term present when the TI surface is gapped and the ordinary 
electromagnetic response. However, as for the present case, the finite frequency behavior 
of the $\theta-$term \cite{GdJ12} can be important at certain length scales. Including such an effect  
for all frequencies is an intricate calculation due to the difficulty of modelling the 
surface exactly. Our results and \eqref{eq:Modelmain}, can serve as a first approximation 
to model the surface since it is gapped and also carries a quantized Hall conductivity.
Interpreting our findings for the particular case where $C_{1,2}=\pm1$ as a zero magnetic field analogue of the results of Ref.~\cite{BV00,TM12} where a QHE system also leads to repulsion has to be understood with caution since (i)
there is a crucial sign difference in \eqref{repulsionlarge} with respect to the QHE case that prevents the simple mapping $C_{i}\to \nu_{i}$, where $\nu_{i}$ is the filling fraction and (ii) the possibility of tuning the sign of the force by simply turning over one of the plates is exclusive to the CI system. Moreover, the refinement and complexity of Casimir experiments makes the disposal of the external magnetic field a particularly valuable feature of the proposed CI system.
The repulsive behavior discussed in this Letter also applies to the fractional version of Chern insulators i.e. fractional Chern insulators (FCI) \cite{NSCC11,TMW11,SGK11} that are many-body incompressible states with Hall conductivities quantized to fractions of $e^2/h$. \\
To conclude, we have shown that two Chern insulator plates with finite 
Chern numbers present a repulsive (attractive)
Casimir effect as long as the Chern numbers have opposing (equal) signs. The force can be tuned to attraction 
by simply turning over one of the plates or by electrostatic doping. 
Our results point towards TI thin films and multi-orbital systems with higher Chern numbers \cite{YGS12,TB12,JQL12,WLZ13,FGB13}
as the most promising future route to realize and control Casimir repulsion.\\
\emph{Acknowledgments:} We thank A. Cortijo, F. de Juan, M. A. H. Vozmediano,
W.-K. Tse for discussions and Diego Dalvit and Frank Pollmann for critical reading of the manuscript.
Support from FIS2011-23713, PIB2010BZ-00512 (A. G. G.) and EPSRC under EP/H049797/1 (P.R.-L) is acknowledged.

\newpage

\appendix

\begin{widetext}

\section{Supplementary Material}

\subsection{Fresnel coefficients for a Chern insulator}

In this section we derive the Fresnel coefficients for a Chern insulator (CI). For the sake of gaining generality,
our starting point is the situation where the CI separates a dielectric medium, characterized with 
the dielectric function and magnetic susceptibility $\varepsilon,\mu$, and the vacuum. 
At the end we will take the limit where the dielectric medium is the vacuum itself ($\varepsilon,\mu\to 1$) which generates
the CI Fresnel coefficients used in the main text. Keeping $\varepsilon,\mu$ finite opens the way to study other phenomena, such as the effect of having the CI placed on a dielectric substrate or 3D topological insulators. \\ 
The electromagnetic response of the CI plate is characterized by a surface current $\mathbf{J}_{S} = \frac{4\pi}{c}\sigma_{s}\mathbf{E}$, where
\begin{eqnarray}
\sigma_{s} = \left(\begin{array}{c|c}
\sigma_{xx} & \sigma_{xy}\\
\hline
\sigma_{yx} & \sigma_{yy}
\end{array}\right) = \left(\begin{array}{c|c}
  \sigma_{xx} & \sigma_{xy}\\
  \hline
- \sigma_{xy} & \sigma_{xx}
\end{array}\right).
\end{eqnarray}
Following the procedure detailed in appendix C of Ref.~\cite{GRC11}, the boundary conditions 
for the electromagnetic fields $\mathbf{n}\times\mathbf{H}=\mathbf{J}_{S}$ and $\mathbf{n}\times\mathbf{E}=0$ 
in the presence of a CI plate 
(we assume that the normal vector is $\hat{\textbf{n}} = \hat{z}$) take the form
\begin{eqnarray}\nonumber
- (E_{die,y} - {E}_{vac,y}) & = & 0,\label{BC_ey}\\
\nonumber
   E_{die,x} - {E}_{vac,x}  & = & 0,\label{BC_ex}\\
- (H_{die,y} - {H}_{vac,y}) & = & \phantom{-} \frac{4\pi}{c}\sigma_{xx}{E}_{die,x} + \frac{4\pi}{c}\sigma_{xy}{E}_{die,y},\label{BC_hy}\\
   H_{die,x} - {H}_{vac,x}  & = &          -  \frac{4\pi}{c}\sigma_{xy}{E}_{die,x} + \frac{4\pi}{c}\sigma_{xx}{E}_{die,y}.\label{BC_hx}
\end{eqnarray}
The incoming and reflected waves can be written as
\begin{eqnarray}\nonumber
\mathbf{E}_{in} &=& \left(A_{\perp}\mathbf{y}+A_{\parallel}\dfrac{c}{\omega}(k_{z}\mathbf{x}-k_{x}\mathbf{z}) \right) e^ {i(k_{x}x+k_{z}z-\omega t)},\\
\nonumber
\mathbf{H}_{in} &=& \left(A_{\parallel}\mathbf{y}-A_{\perp}\dfrac{c}{\omega}(k_{z}\mathbf{x}-k_{x}\mathbf{z}) \right) e^ {i(k_{x}x+k_{z}z-\omega t)},
\end{eqnarray}
and
\begin{eqnarray}\nonumber
\mathbf{E}_{r} &=& \left(R_{\perp}\mathbf{y}-R_{\parallel}\dfrac{c}{\omega}(k_{z}\mathbf{x}+k_{x}\mathbf{z}) \right) e^ {i(k_{z}x-k_{x}z-\omega t)},\\
\nonumber
\mathbf{H}_{r} &=& \left(R_{\parallel}\mathbf{y}+R_{\perp}\dfrac{c}{\omega}(k_{z}\mathbf{x}+k_{x}\mathbf{z}) \right) e^ {i(k_{x}x-k_{z}z-\omega t)},
\end{eqnarray}
respectively, where we have used that $k_{x}^{in}=-k_{x}^{ref}$. We define $k_{x}=\frac{\omega}{c}\mathrm{sin}\theta_{i}$ and $k_{z}=\frac{\omega}{c}\mathrm{cos}\theta_{i}$, where $\theta_{i}$ is the angle of incidence. The quotients between the relative amplitudes $A_{\perp},A_{\parallel}$ and $R_{\perp},R_{\parallel}$ will define the entries of the reflection matrix.
Following the same steps as in appendix C in \cite{GRC11} we can write the boundary conditions as
\begin{eqnarray}
  (A_{\parallel} - R_{\parallel})\frac{c}{\omega}k_{z} & = & e_{x},\\
   A_{\perp} + R_{\perp} & = & e_{y},\\
- (A_{\parallel} + R_{\parallel}) & = & - h_{y} + \frac{4\pi}{c}\sigma_{xx}e_{x} + \frac{4\pi}{c}\sigma_{xy}e_{y},\\
  (R_{\perp} - A_{\perp})\frac{c}{\omega}k_{z} & = & \phantom{-} h_{x} - \frac{4\pi}{c}\sigma_{xy}e_{x} + \frac{4\pi}{c}\sigma_{xx}e_{y},
\end{eqnarray}
which leads to the reflection matrix 
\begin{eqnarray}\label{Definition_Rmatrix}
\mathbb{R} = \dfrac{1}{\Delta}\left(
\begin{array}{cc}
\tilde{r}_{ss} & \tilde{r}_{sp}\\
\tilde{r}_{ps} & \tilde{r}_{pp}
\end{array}\right),
\end{eqnarray}
with
\begin{eqnarray}
\tilde{r}_{ss} & = & c^{2}(\mu k_{z} - q) \left[ 4\pi q\,k_{z}\sigma_{xx} + \omega\left(\epsilon\,k_{z} + q\right) \right] - 4\pi\mu\omega\left[ \omega\sigma_{xx}\left(\epsilon\,k_{z} + q\right) + 4\pi q\,k_{z}\left(\sigma_{xx}^{2} + \sigma_{xy}^{2} \right)\right],\\
\tilde{r}_{sp} & = & 8\pi\,c\,q\,k_{z}\mu\omega\sigma_{xy},\\
\tilde{r}_{ps} & = & 8\pi\,c\,q\,k_{z}\mu\omega\sigma_{xy},\\
\tilde{r}_{pp} & = & c^{2}(\mu k_{z} + q) \left[ 4\pi q\,k_{z}\sigma_{xx} + \omega\left(\epsilon\,k_{z} - q\right) \right] + 4 \pi\mu\omega\left[ \omega\sigma_{xx}\left(\epsilon\,k_{z} - q\right) + 4\pi q\,k_{z} \left(\sigma_{xx}^{2} + \sigma_{xy}^{2} \right)\right],\\
\Delta & = & c^{2} (\mu k_{z} + q)\left[ 4\pi q\,k_{z}\sigma_{xx} + \omega\left(\epsilon\,k_{z} + q\right) \right] + 4\pi\mu\omega\left[ \omega\sigma_{xx}\left(\epsilon\,k_{z} + q\right) + 4\pi q\,k_{z}\left(\sigma_{xx}^{2} + \sigma_{xy}^{2} \right)\right],
\end{eqnarray}
where $k^2_{z}=\frac{\omega^2}{c^2}-k_{\parallel}^2$ and $q^2=\frac{\omega^2}{c}\epsilon\mu-k_{\parallel}^2$.\\
From this result, it is interesting to note that we recover the reflection matrix of a three dimensional topological insulator used 
in \cite{GC11} in the limit where $\sigma_{xx}\to\,0$ and defining $\frac{4\pi}{c}\sigma_{xy}\equiv\bar{\alpha}$. This is a manifestation of the fact that a 3D TI described by a $\theta-$term is a dielectric with a ``Hall effect" at the surface. These are therefore the generalization of the TI coefficients in \cite{GC11} when the surface has not only a Hall but a longitudinal conductivity.  As mentioned in the main text, this result also can be regarded as the first step to study the consequences of finite frequency axionic response of a Topological insulator (calculated in Ref.~\cite{GdJ12}) in the context of the Casimir effect.
As outlined above, in order to obtain the Fresnel coefficients for a CI plate we set $\varepsilon,\mu\to 1$ in \eqref{Definition_Rmatrix}, arriving to
\begin{eqnarray}
\tilde{r}_{ss} & = & - 2\pi\left( \frac{\sigma_{xx}}{c\lambda} + \dfrac{2\pi}{c^2}\left( \sigma_{xx}^{2} + \sigma_{xy}^{2} \right)\right),\\
\tilde{r}_{sp} & = & 2 \pi\sigma_{xy}/c,\\
\tilde{r}_{ps} & = & 2 \pi\sigma_{xy}/c,\\
\tilde{r}_{pp} & = & 2 \pi\left( \lambda\dfrac{\sigma_{xx}}{c} + \dfrac{2\pi}{c^2}\left( \sigma_{xx}^{2} + \sigma_{xy}^{2} \right)\right),\\
\Delta & = & 1 + 2\pi\dfrac{\sigma_{xx}}{c} \left( \frac{1}{\lambda} + \lambda \right) + \dfrac{4\pi^{2}}{c^2}\left( \sigma_{xx}^{2} + \sigma_{xy}^{2} \right),
\end{eqnarray}
used in the main text (see also Ref.~\cite{TM12}).
Here $k_{z}^{2} = \frac{\omega^{2}}{c^{2}} - k^{2}_{\parallel} = q^{2}$ and $\lambda = \frac{k_{z}c}{\omega}$. 

\section{Chern Insulator model and optical conductivity from the Kubo formula}
In the main text we have made use of the generic model for CI given by
\begin{eqnarray}
\label{Modelapp}
&&
H^{\,}_{0}=
\sum_{\bs{k}\in\mathrm{BZ}}
c^{\dagger}_{\bs{k}}
\,\bs{d}^{\,}_{\bs{k}}\cdot\bs{\sigma}\,
c^{\,}_{\bs{k}},
\\
&&
d^{\,}_{\bs{k};1}+\mathrm{i}\,d^{\,}_{\bs{k};2}=
t(\sin\,k^{\,}_{1}+\mathrm{i}\,\sin \, k^{\,}_{2}),
\\
&&
d^{\,}_{\bs{k};3}=
h^{\,}_{1}\cos\,k^{\,}_{1}
+h^{\,}_{2}\cos \, k^{\,}_{2}
+h^{\,}_{3}
\nonumber
\\
&&
\hphantom{d^{\,}_{\bs{k};3}:=}
+
h^{\,}_{4}
\left[
\cos(k^{\,}_{1}+k^{\,}_{2})+\cos(k^{\,}_{1}-k^{\,}_{2})
\right],
\nonumber\\
&&
\end{eqnarray}
%\end{subequations}
where  $
c^{\dag}_{\bs{k}}\equiv
(c^{\dag}_{\bs{k},\uparrow},c^{\dag}_{\bs{k},\downarrow})
$ and $c^{\dag}_{\bs{k},s}$ creates a fermion at momentum $\bs{k}$ in the Brillouin zone 
(BZ) with spin $s=\uparrow, \downarrow$ while $\bs{\sigma}=(\sigma^{\,}_{1},\sigma^{\,}_{2},\sigma^{\,}_{3})$ 
are the three Pauli matrices acting on spin space. The parameters $t$ and $h^{\,}_\mu,\ \mu=1,\cdots,4$, are real
and effectively are used to tune the Chern number of the system.
As explained in Ref.~\cite{GNC12} the low energy Hamiltonian around the four
inversion-symmetric points $\bs{k}^{(ij)}=\pi(i,j),\ i,j=0,1$ to linear order produces four low
energy gapped Dirac Hamiltonians with masses given by
\label{eq: noninteracting H momentum space}

\begin{equation}
m^{(ij)}=
(-1)^i h^{\,}_{1}
+(-1)^j h^{\,}_{2}
+h^{\,}_{3}
+(-1)^{i+j}\,2\,h^{\,}_{4}.
\label{eq: massterms}
\end{equation}
The Chern number of each of the two bands is well defined whenever the system is gapped. 
Each Dirac point contributes $\pm1/2$ to the Chern number, 
depending essentially on the sign of the masses at each cone.
The total Chern number of the lower band can be written as
\begin{equation}\label{eq: Cnumber}
C=\frac{1}{2}\sum_{i,j=0,1} (-1)^{i+j}\,\mathrm{sgn}\,m^{(ij)},
\end{equation}
and therefore it can span the values $C=\{0,\pm1,\pm2\}$.\\
For a Dirac model like \eqref{Modelapp},
of the form $H_{\bs{k}}=d_{i,\bs{k}}\sigma_{i}+\epsilon_{\bs{k}}\mathsf{1}$
the Kubo formula for the optical conductivity $\sigma_{ij}(\omega)$ 
takes the form
\begin{eqnarray}\label{Kubo1}
\sigma_{ij}(\omega)&=&\dfrac{i}{\omega}K_{ij}(\omega+i\delta),\\
\label{Kubo2}
K_{ij}(\nu_{m})&=&\dfrac{1}{\Omega N}\dfrac{1}{\beta}\sum_{\omega_{n},\bs{k}}
\mathrm{Tr}G_{\omega_{n},\bs{k}}J^{i}_{\bs{k}}G_{\omega_{n}+\nu_{m},\bs{k}}J^{j}_{\bs{k}},
\end{eqnarray}
where $\beta=1/k_{B}T$, $N$ is the number of unit cells of volume $\Omega$, $J_{\omega_{n},\bs{k}}$ is the current operator defined by
$J^{i}_{\bs{k}}=\dfrac{\partial H_{\bs{k}}}{\partial k_{i}}$ and 
and $G_{\omega_{n},\bs{k}}$ is the Matsubara Green's function. The latter can be written as
\begin{eqnarray}
G_{\omega_{n},\bs{k}}=(i\omega_{n}-H_{\bs{k}})^{-1}=\sum_{s=\pm}\dfrac{P_{s,\bs{k}}}{i\omega_{n}-E_{s,\bs{k}}},
\end{eqnarray}
with $E_{\pm,\bs{k}}=\pm|d_{\bs{k}}|+\epsilon_{\bs{k}}$ and $P_{s,\bs{k}}=\dfrac{1}{2}(1\pm \sum_{i}d_{i,\bs{k}}\sigma_{i})$. Introducing
this last equation in \eqref{Kubo2} 
\begin{eqnarray}\label{Kubo3}
K_{ij}(i\nu_{m})&=&\dfrac{1}{\Omega N}\sum_{s,t=\pm}\sum_{\bs{k}}
\dfrac{\mathrm{Tr}\left[J^{i}_{\bs{k}}P_{s,\bs{k}}J^{j}_{\bs{k}}P_{t,\bs{k}}\right]}{i\nu_{m}-E_{s,\bs{k}}+E_{t,\bs{k}}}(n_{t,\bs{k}}-n_{s,\bs{k}}),
\end{eqnarray}
where $n_{t,\bs{k}}=(e^{\beta(E_{t,\bs{k}}-\mu)}+1)^{-1}$ are the Fermi distribution functions and we assume for all cases that the chemical potential $\mu$ is inside the single particle gap.\\
 The usual change
 $i\nu_{m}\to\omega+i\delta$ leads to the final expression for $\sigma_{ij}(\omega)$ for the Dirac Hamiltonian $H_{\bs{k}}=d_{i,\bs{k}}\sigma_{i}+\epsilon_{\bs{k}}\mathsf{1}$. This expression can also be evaluated at $\omega=i\xi$ to obtain the conductivity in the imaginary axis as well, required for the evaluation of the Casimir energy density \eqref{CasimirEnergy}. \\
In Fig. \ref{fig: condmain} of the main text we present a typical example of the band structure, longitudinal and Hall conductivities calculated with \eqref{Kubo3} 
for $\beta=10^4$ and  $h_\mu=(1,1,0.25,0)t$ that corresponds to a case 
with lower band of $C=1$. We also show 
the analytical result for a massive Dirac Hamiltonian 
with $m$ up to $\omega\sim 2m$ \cite{HSZ11}
\begin{eqnarray}\label{eq:analyticcond1}
\mathrm{Re}\sigma_{xx}(\omega)&=&\dfrac{e^2}{h}\dfrac{\pi}{4}\left(1+\dfrac{4m^2}{\omega^2}\right)\Theta(\omega^2-4m^2),\\
\label{eq:analyticcond2}
\mathrm{Re}\sigma_{xy}(\omega)&=&\dfrac{e^2}{h}\dfrac{m}{\omega}\mathrm{ln}\left|\dfrac{2m+\omega}{2m-\omega}\right|.
\end{eqnarray}
which agrees with the numerical calculation for low energies up to $\omega\sim 2m$.
% 
%\begin{figure}
%\includegraphics[angle=0,scale=0.2,page=3]{cond.pdf}
%\caption{\label{fig: cond}
%(Color online) a) Band structure b) $\sigma_{xy}$ and c) $\sigma_{xx}$ as a function of real and imaginary frequencies for the model \eqref{Modelapp} calculated with \eqref{Kubo3} for $\beta=10^4$, and $h_\mu=(1,1,0.25,0)t$. The conductivities are given in units of $e^2/h$. A comparison is shown between the numeric and the analytical formulas \eqref{eq:analyticcond1} and \eqref{eq:analyticcond2} . }
%\end{figure}
%%
Other cases, present qualitatively the same behaviour, with quantization of the Hall conductivity at $Ce^2/h$. 
To calculate $\sigma_{ij}(\omega=i\xi)$ it is possible to integrate directly the Kubo formula following for example Ref.~\cite{DPW12}
or use the Kramers-Kronig (KK) dispersion relations as provided in the next section.
\section{Kramers-Kronig dispersion relations}
In this section we review the procedure to calculate $\sigma_{ij}(i\xi)$ from $\sigma_{ij}(\omega)$ by using the Kramers-Kronig (KK) relations. The starting point are the KK relations for the dielectric function \cite{LLv8}
\begin{eqnarray}\label{Reeps}
\mathrm{Re}\left[\varepsilon(\omega)-1\right]=\dfrac{2}{\pi}\mathcal{P}\int_{0}^{\infty}d\tilde{\omega}\dfrac{\tilde{\omega}\mathrm{Im}\varepsilon(\tilde{\omega})}{\tilde{\omega}^2-\omega^2}
%\label{Imeps}
%\mathrm{Im}\left[\varepsilon(\omega)\right]=-\dfrac{2}{\pi}\omega\int_{0}^{\infty}d\tilde{\omega}\dfrac{\mathrm{Re}\varepsilon(\tilde{\omega})}{\tilde{\omega}^2-\omega^2}
\end{eqnarray}
The relation to the conductivity $\sigma(\omega)=\sigma_{1}(\omega)+i\sigma_{2}(\omega)$ is given by (see for example Ref.~\cite{DG02} Table 2.1)
\begin{eqnarray}\label{epscond}
\varepsilon(\omega)=1+4\pi i\dfrac{\sigma(\omega)}{\omega}
\end{eqnarray}
Evaluating the real part of $\varepsilon(\omega)$ for $\omega\to i\xi$
\begin{eqnarray}\nonumber
\mathrm{Re}\left[\varepsilon(i\xi)-1\right]&=&\dfrac{2}{\pi}\int_{0}^{\infty}d\tilde{\omega}\dfrac{\tilde{\omega}\mathrm{Im}\left[4\pi i\dfrac{\sigma(\tilde{\omega})}{\tilde{\omega}}\right]}{\tilde{\omega}^2+\xi^2}\\
\nonumber
\mathrm{Re}\left[ \dfrac{\sigma(i\xi)}{\xi}\right]&=&\dfrac{2}{\pi}\int_{0}^{\infty}d\tilde{\omega}\dfrac{\sigma_1(\tilde{\omega})}{\tilde{\omega}^2+\xi^2}\\
\label{Recondimgood}
\sigma(i\xi)&=&\dfrac{2}{\pi}\int_{0}^{\infty}d\tilde{\omega}\dfrac{\xi\sigma_1(\tilde{\omega})}{\tilde{\omega}^2+\xi^2}
\end{eqnarray}
We have used that the imaginary part of $\sigma(i\xi)$at the imaginary axis is zero \cite{LLv8}.\\
It is possible to check that this results indeed coincides with the Kubo formula evaluated at $\omega\to i\xi$. 
For example,
taking a massive (gapped) Dirac Fermion as an example, the longitudinal conductivity is given by
\begin{equation}\label{sxxDF}
\sigma_{xx}(\omega)=\dfrac{e^2}{h}\dfrac{\pi}{4}\left(1+\dfrac{4m^2}{\omega^2}\right)\Theta(\omega^2-4m^2)
\end{equation}
where $m$ is the gap. Using \eqref{Recondimgood} the integral to perform is
\begin{eqnarray}
\label{RecondimgoodDF}
\sigma(i\xi)&=& \dfrac{2}{\pi}\xi\int_{2m}^{\infty}d\tilde{\omega}\dfrac{e^2}{h}\dfrac{\pi}{4}\left(1+\dfrac{4m^2}{\omega^2}\right)\dfrac{1}{\tilde{\omega}^2+\xi^2}
\end{eqnarray}
This coincides with the Kubo formula evaluated at imaginary frequencies (see for example eq. (7) of \cite{DPW12} at zero temperature and chemical potential. In this limit the intraband contribution of eq. (6) in \cite{DPW12} vanishes).

\subsection{Position of $d_{max}$ and power law scaling of the Casimir energy density}

In this section we derive analytically the power law governing the long and short distance 
limits of the Casimir energy density for two Chern insulator plates finally giving additional
numerical support for this calculation. In doing so, we also estimate the position of $d_{max}$.\\

\textbf{Long distance limit:} Here we will follow the recipe of Ref.~\cite{TM12}.
The long distance limit of the Casimir energy density 
is given by the behavior of the Hall and longitudinal conductivities 
of the Chern Insulator plates at low frequency, which to order $\omega$ are
\begin{eqnarray}
\sigma_{xx,i}(i\omega)&\simeq & \dfrac{d\sigma_{xx}(i\omega)}{d\omega}\bigg\vert_{\omega=0}\omega\simeq b_{i} \alpha \omega,\\
\sigma_{xy,i}(i\omega)&\simeq & \to C_{i}\frac{\alpha}{2\pi},
\end{eqnarray}
where $C_{i}$ is the Chern number characterizing each plate, $b_{i}$ is the first
coefficient of a Taylor expansion and $\alpha=e^2/(\hbar c)$ is the fine structure constant. 
The reflection matrix for a CI plate in this approximation
is given by
\begin{eqnarray}\label{Definition_Rmatrixb}
\mathbb{R}_{i} = \left(
\begin{array}{c|c}
- \alpha^{2} C_{i}^{2} - 2\pi b_{i}\kappa\alpha\lambda^{-1} & C_{i}\alpha - 2\pi C_{i}b_{i}\kappa\alpha^{2}\left( \lambda + \lambda^{-1} \right)\\
\hline
C_{i}\alpha - 2\pi C_{i}b_{i}\kappa\alpha^{2}\left( \lambda + \lambda^{-1} \right) & \alpha^{2} C_{i}^{2} + 2\pi b_{i}\kappa\alpha\lambda
\end{array}\right),
\end{eqnarray}
where $\lambda = \frac{1}{\kappa}\sqrt{ \kappa^{2} + \textbf{k}_{\parallel}^{2}}$, $\kappa=\xi/c$.
Therefore, the long distance limit depends on the number of Chern Insulators with $C_{i}\neq 0$.
Carrying out explicitly the integral \eqref{CasimirEnergy}  to order $\alpha^2 $ one arrives at
\begin{eqnarray}\label{Large_Distance_Approximation}
E & = & - \frac{\hbar c \alpha^{2}}{8\pi^{2} d^{3}}C_{1} C_{2}- \frac{\hbar c\,\alpha^{3}}{4\pi d^{4}}\left(C_{1}^{2}b_{2} + C_{2}^{2}b_{1} - 2C_{1}C_{2}(b_{1} + b_{2})\right)- \frac{9\hbar c\,\alpha^{2}}{10\,d^{5}}b_{1}b_{2}.
\end{eqnarray}
The first term, of order $\alpha^2$ is eq.~\eqref{repulsionlarge} of the main text which decreases as $\sim 1/d^{3}$. If both of the Chern numbers are zero the power law changes to $\sim 1/d^{5}$ governed by the third term, also of order $\alpha^2$. Thus, an estimate for the length scale $d_{max}$ is given by the crossover from the first to the third term. A straight forward calculation leads to the power law $d_{max}\sim 1/\sqrt{|C_{1}C_{2}|m_{1}m_{2}}$ since $b_{i}\sim1/m_{i}$ and $m_{i}$ is the single particle mass gap, the main scale of the problem measured in units of $t$. If only one of the Chern numbers $C_{i}$ is zero, it is necessary to consider the next term in the expansion, of order $\alpha^3$ that decays as $\sim1/d^{4}$. 
\\
To complement this analysis we have also studied the behaviour of the Casimir Energy density at large
Chern numbers. The complete asymptotic formula for the Casimir energy between non-zero Chern Insulators is
\begin{equation}\label{Complete_Asymptotic_CIEnergy}
\lim_{d\to\infty}E_{12} = - \frac{\hbar c \alpha^{2}}{8\pi^{2} d^{3}}\frac{C_{1} C_{2} \left(1 + C_{1} C_{2} \alpha^{2}\right)}{\left(1 + C_{1}^{2} \alpha^{2}\right) \left(1 + C_{2}^{2} \alpha^{2}\right)},
\end{equation}
which, in the limit $\alpha \ll 1$ recovers \eqref{Large_Distance_Approximation}.
In what follows we will focus in two particular cases, (i) when $C_{1} = C_{2}\equiv C_{+}$ 
and when $C_{1} = - C_{2}\equiv C_{-}$.
It is instructive to notice that for large Chern numbers and when $|C_{1}|=|C_{2}|\equiv C$
the system is always attractive \emph{irrespective of the sign of the Chern numbers}.
In this extreme case the Casimir energy tends to
\begin{equation}
\lim_{C \gg 1}\lim_{d\to\infty}E_{12} =- \frac{\hbar c }{8\pi^{2}d^{3}} \approx 0.92 E_{PM},
\end{equation}
which is a Casimir energy of the same magnitude than the metal-metal case $E_{PM}=-\hbar c\pi^2/(720d^3)$. As discussed in the text,
this limit is completely inaccessible experimentally. Note that therefore, when $C_{1}$ and $C_{2}$ are equal but with opposite sign ($C_{1} = - C_{2}=C_{-}$), the system is not always repulsive at large distances. The maximum repulsive Casimir energy is reached when $C_{-} = \frac{1}{\sqrt{3}\alpha }$, with magnitude approximately a tenth part of the perfect metal Casimir energy. The system becomes attractive when $C_{-} > \frac{1}{\alpha} \approx 137$ reaching the same asymptotic result as with $C_{1}=C_{2}$.
These results are summarized in Fig.~\ref{fig: mass dependence} b).
\\

\textbf{Short distance limit:} The short distance limit of the Casimir energy density
is given by the behavior of the conductivities of the CI at high frequencies.
For large frequencies the conductivity tensor behaves as
\begin{eqnarray}
\sigma_{xx}^{(i)}(i\omega)&\simeq &\alpha\frac{s_{xx,i}}{\omega} \ll 1,\\
\sigma_{xy}^{(i)}(i\omega)&\simeq &C_{i}\alpha\frac{s_{xy,i}}{\omega^{2}} \ll 1,
\end{eqnarray}
and the reflection matrix is given by
\begin{equation}\label{Definition_Rmatrix_2}
\mathbb{R} = 
\left(
\begin{array}{cc}
- \frac{2\pi  s_{xx}\alpha}{4\pi s_{xx}\alpha + \kappa\lambda} & 0 \\
 0 & \frac{2\pi  s_{xx}\alpha\lambda}{2\pi s_{xx}\alpha\left( \lambda + \lambda^{-1} \right) + \kappa}
\end{array}
\right),
\end{equation}
which scales with $\alpha$, i.e. $\|\mathbb{R}\|  \ll 1$ independent of $C_{i}$. 
Then, we can calculate analytically the Casimir energy density in this limit by the use of the approximation
\begin{equation}
\log\mathrm{det}\left[ \mathbb{I} - \mathbb{N}\right] \approx - \mathrm{Tr}\left[\mathbb{N}\right],
\end{equation}
in \eqref{CasimirEnergy}. 
After carrying out the integrals, we obtain that the Casimir energy density is
\begin{equation}\label{Short_Distance_Approximation}
E_{0} = -\frac{3\hbar c}{128}\sqrt{\frac{\alpha}{d^{5}}}\frac{\sqrt{s_{xx,1}s_{xx,2}}}{\sqrt{s_{xx,1}}+\sqrt{s_{xx,2}}},
\end{equation}
which implies a power law behavior of $\sim1/d^{5/2}$ at short distances regardless of the Chern number. \\

\textbf{Numerical results:} To conclude this section we provide numerical evidence for the above power law behaviours. 
Fig.~\ref{fig: d-dependence} shows the absolute value of the Casimir energy density between Chern Insulators  
as a function of the distance $d$ for two distinct cases where
(a) the two Chern numbers are finite and (b) one (or both) of the Chern numbers is (are) zero.
To compare with the different asymptotic results, in each figure we provide dashed lines that scale as predicted by 
\eqref{Large_Distance_Approximation}  and \eqref{Short_Distance_Approximation}. In agreement with these, we observe that 
for short distances, the force is attractive and proportional to $d^{-5/2}$ independently of $C_{1}$ and $C_{2}$ given that they are non-zero.
At large distances the energy scales as $d^{-3}$ for finite values of $C_{1}$ and $C_{2}$ and as $d^{-4}$ ($d^{-5}$) if one (both) of them is (are) zero in agreement with the above analytical arguments. We emphasize that the agreement of both the long and short distance limits is both quantitative and qualitative although we chose to represent the analytical result in Fig.~\ref{fig: d-dependence} with a slight offset for clarity.
\begin{figure}
\includegraphics[angle=0,scale=0.2,page=4]{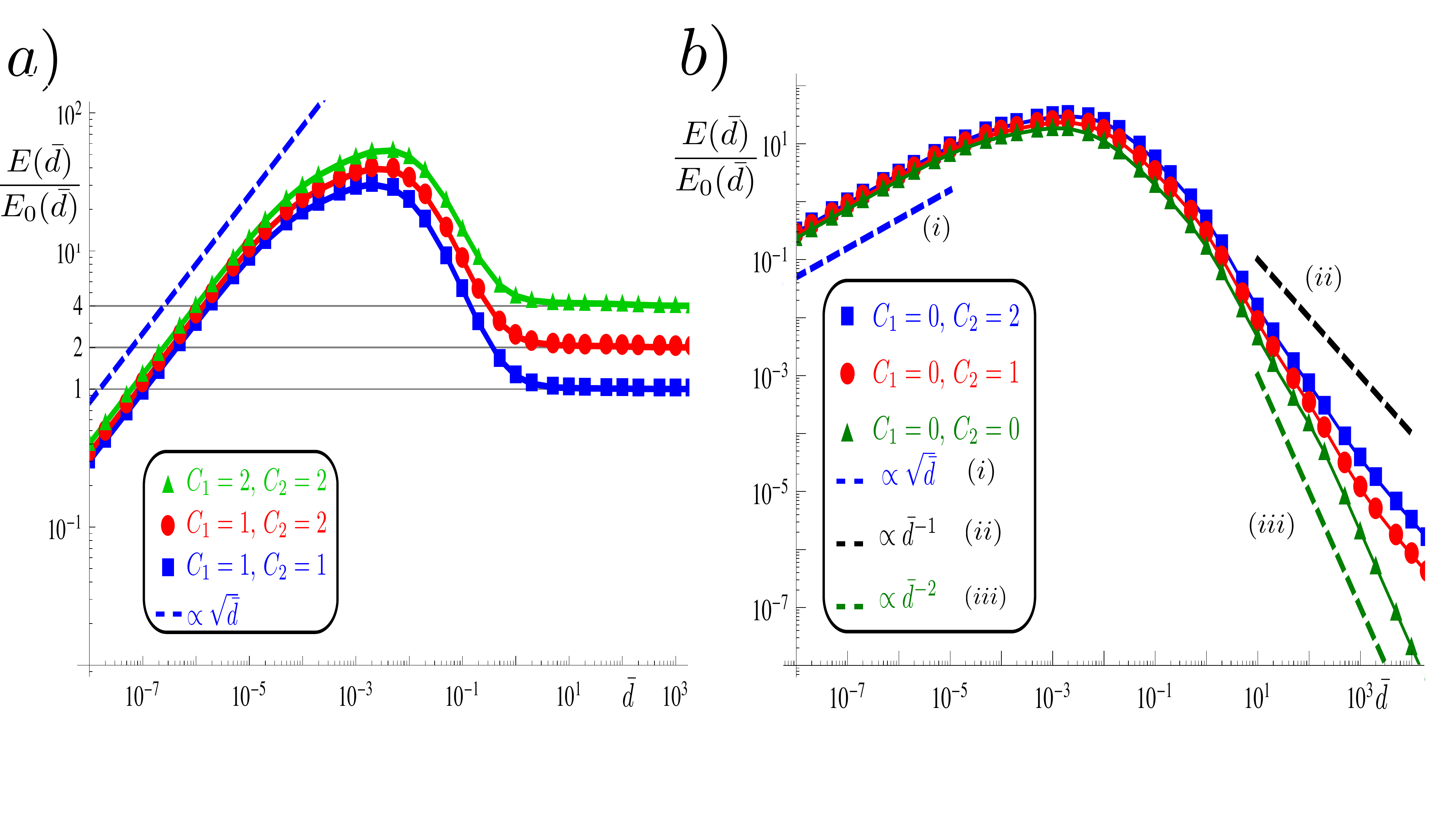}
\caption{\label{fig: d-dependence}
(Color online) Absolute value of Casimir energy density in units of $E_{0}(\bar{d})=\hbar c \alpha^{2}/(8\pi^2\bar{d}^{3})$ between 
CI plates as a function of the dimensionless distance $\bar{d}$ for the cases with (a) $C_{i}\neq 0$ for both plates and 
(b) $C_{i}=0$ for either or both $i=1,2$ in a log-log scale. The different decay power laws are shown 
as dashed lines with a slight offset for clarity although the agreement with the analytical results 
when the rescaling factor $E_{0}(\bar{d})$ is taken into account is both qualitative and quantitative.}
\end{figure}

\end{widetext}

\end{document}